\documentclass[12pt,epsfig]{article}

\usepackage{amssymb,epsfig}

\newcommand{\be}{\begin{equation}}

\newcommand{\ee}{\end{equation}}

\newcommand{\bea}{\begin{eqnarray}}

\newcommand{\eea}{\end{eqnarray}}

\newcommand{\kt}{\vec{k}}
\newcommand{\p}{\vec{p}}

\newcommand{\2}{\vec{p}_{2\pi}}

\begin{document}

\begin{titlepage}

\begin{flushright}
\begin{tabular}{l}
 CPHT S 001.01.02\\
 hep-ph/0202231
\end{tabular}
\end{flushright}
\vspace{1.5cm}

\begin{center}

{\LARGE \bf  Hunting the QCD-Odderon in
hard diffractive electroproduction of two pions}
\vspace{1cm}

{\sc Ph.~H{\"a}gler}${}^1$,
{\sc B.~Pire}${}^{2}$,
{\sc L.~Szymanowski}${}^{2,3}$ and
{\sc O.V.~Teryaev}${}^{4}$
\\[0.5cm]
\vspace*{0.1cm} ${}^1${\it
   Institut f{\"u}r Theoretische Physik, Universit{\"a}t
   Regensburg, \\ D-93040 Regensburg, Germany
                       } \\[0.2cm]
\vspace*{0.1cm} ${}^2$ {\it
CPhT, {\'E}cole Polytechnique, F-91128 Palaiseau, France\footnote{
  Unit{\'e} mixte C7644 du CNRS.}
                       } \\[0.2cm]
\vspace*{0.1cm} ${}^3$ {\it
 So{\l}tan Institute for Nuclear Studies,
Ho\.za 69,\\ 00-681 Warsaw, Poland
                       } \\[0.2cm]
\vspace*{0.1cm} ${}^4$ {\it
Bogoliubov Lab. of Theoretical Physics, JINR, 141980 Dubna, Russia
                       } \\[1.0cm]
{\it \large
\today
 }
\vskip2cm
{\bf Abstract:\\[10pt]} \parbox[t]{\textwidth}{
  Charge asymmetries in diffractive electroproduction of two mesons are
proportional to the interference of Pomeron and Odderon exchange
amplitudes. We calculate in the framework of QCD and in the Born
approximation a forward-backward charge asymmetry which turns out to be
sizable
in a kinematical domain accessible to HERA experiments. We predict a
distinctive dependence of this asymmetry on the invariant mass of the two
pions. Testing this prediction is a crucial step in the discovery of the
QCD-Odderon. }
\vskip1cm
\end{center}

\vspace*{1cm}

\end{titlepage}


{\large \bf 1.~~}
Pomeron and Odderon  exchanges are the theoretically dominant
contributions to  hadronic cross sections at high energy. They appear
on an equal footing in the QCD description of hadronic reactions, and
in the lowest order approximation they
correspond to  colour singlet exchanges in the $t$-channel with
 two and three gluons, respectively.

\vskip.1in
The relevance of the Odderon exchange for hadronic reactions was
emphasized long ago
\cite{LN}. In perturbative QCD the Odderon is described by the
Bartels-Kwiecinski-Praszalowicz (BKP) equation \cite{BKP}.
In spite of many attempts to solve the BKP equation, its
solutions are still known only partially although much progress has been
recently achieved
 \cite{Levodd, JW, Vacca1, Korch}. Still this appears to be in a
strong contrast
 with the QCD description of the Pomeron, where  solutions  of
 the leading order  Balitsky-Fadin-Kuraev-Lipatov  (BFKL) equation
are well known \cite{BFKL}.

\vskip.1in
The Odderon remains a mistery also from an experimental point of view.
On the one hand, recent studies of the elastic $pp$ scattering
show that one needs the Odderon contribution to understand the
data in the
dip region \cite{Doshrecent}.
On the other hand, the studies of reactions which should select only the
odderon exchange didn't show any clear signal of its importance.
In the case of  diffractive $\eta_c$-meson photoproduction,
the  QCD prediction in the Born approximation for the cross section
is rather small \cite{KM,Engel}.
The inclusion of evolution following from the BKP equation
\cite{Vacca2} leads to an increase of the predicted
cross section for this process by  one order of magnitude.
Recent experimental studies at HERA of 
exclusive $\pi^0$ photoproduction  \cite{Olsson}  also indicate a very
small cross section for this process, in disagreement with  theoretical
predictions based on the stochastic vacuum model \cite{Dosh}.
In all these meson production processes the scattering amplitude
describing Odderon
exchange enters quadratically in the cross section.

\vskip.1in
In \cite{Brodsky} it was suggested to
study Odderon effects at the amplitude level by means of the
asymmetries in  open charm production.  Since the final state
quark-antiquark pair has no  definite charge parity    both
Pomeron and Odderon exchanges contribute to this process.
The Odderon amplitude enters linearly in the asymmetries and therefore
one can hope  that Odderon effects can show up  more easily.
Moreover, the difficulty with the understanding of soft processes in
QCD  calls for studies of Odderon
contributions in hard processes, such as
electroproduction, where factorization properties allow to perturbatively 
calculate a short-distance part of the scattering amplitude.

\vskip.1in
In this paper, we propose to take
advantage of a number of interesting features of the two pion diffractive
electroproduction process to search for the QCD-Odderon at the amplitude
level.
Here again the two pion state doesn't have
 any definite charge parity and  both Pomeron and Odderon
exchanges contribute. The authors of  Ref. \cite{Nikolaev} 
suggest to study  the charge asymmetry
in soft photoproduction of two pions to select the
interference of the two amplitudes.
Our work shares a number of features with this work.
The originality of our study of the electroproduction process
 is to work
in a perturbative
QCD framework which enables us to derive well founded predictions in an
accessible kinematical domain.


\vskip.1in
The aim of  the present paper is to study the
charge  asymmetries in the reaction
\be
\label{ep}
e^-(p_e)\;\; p(p_N) \to e^-(p_e')\;\; \pi^+(p_+)\;\; \pi^-(p_-)\;\;
p'(p_N')
\ee
 within perturbative QCD, see Fig. 1.  
This
includes the convolution of perturbatively calculable hard subprocess 
with the two non-perturbative inputs~: 2-pion generalized distribution
amplitude (GDA) and Pomeron-Odderon (P/O) proton impact factors. 
GDA \cite{DGPT} are just
the light-cone wave functions of the two
pion system.
Contrary to, say, $\rho$-meson wave functions, they do not require
the selection of a particular charge parity and so are ideally suited for
 studies of the P/O interference.
In order to justify the use of perturbation theory for this
process we consider the electroproduction of this system in which the
hard scale is supplied by the squared mass $-Q^2$
of the virtual photon, $Q^2$ being of the order of a few GeV$^2$.

%
\begin{figure}[t]
\centerline{\epsfxsize12.0cm\epsffile{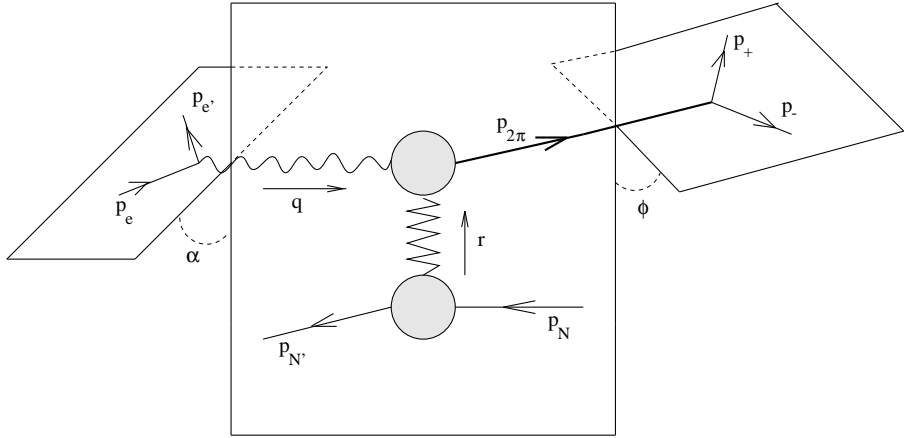}}
\caption[]{\small
Kinematics of the electroproduction of two pions
 }
\label{fig:1}
\end{figure}
%

\vskip.1in

We intend to study the dominant (for asymptotically large $Q^2$)
contribution
to the asymmetries. 
This corresponds to the process with longitudinally
polarized photon. 
In this case the "longitudinal part" of the two pion wave
function is selected,  which is a
straightforward  generalization of the longitudinal polarization of
vector meson.  The longitudinal polarization vector is enhanced by a
Lorentz boost, together with the  cross-section of longitudinally
polarized pion pair production.
This is the twist-2 contribution of the subprocess
$\gamma^* \mbox{P/O} \to q\bar q \to \pi^+ \pi^-$, i.e. in the case of
the
collinear
factorization for the $q\bar q \to \pi^+\, \pi^-$ transition.
At moderate $Q^2$ it is customary to take into account the
transverse momentum $k_\perp$ 
dependence of the meson wave functions which is a model way of accounting
for higher twist contributions, so that for growing $Q^2$ the result is
undistinguishible from the one of standard collinear factorization
\footnote{We postpone the consideration of $k_\perp$-dependence for
future work.}.
The relation between the initial and final polarization states is due 
to the fact that for high-energy diffractive processes 
s-channel helicity conservation is quite well satisfied.
Since the
cross sections $\sigma_L$ and
$\sigma_T$ for the longitudinal and transverse photon polarization,
respectively, are of the same order of magnitude at moderate $Q^2$
\cite{sigmaLT}, an experimental separation of $\sigma_L$ is highly
desirable before confronting our predictions with data.

\vskip.1in

Since the transverse polarization of the pion pair is the only source of
the
amplitude dependence
on the azimuthal angle of the pions in their c.m. frame,
the amplitudes and cross sections are independent of this
angle in our approximation. As a result, the transverse charge
asymmetry, resulting from the distribution in this angle and
discussed in \cite{Nikolaev}, is zero.
Due to that restriction, we  only
study the forward-backward charge asymmetry.

\vskip.1in
In the present study we calculate the lowest perturbative
order contribution to the charge asymmetry, i.e. without taking into
account  the evolution following from the BFKL or the BKP equation. Our
results
should be therefore treated as an estimate of the asymmetries.
The above mentioned  evolutions can be included into the
scattering amplitudes
in a similar way as in Ref. \cite{Vacca2}.


\vskip.2in
{\large \bf 2.~~}
The basic object necessary to calculate the charge asymmetry is the
scattering amplitude for the process with a longitudinal virtual
photon (Fig. 2). 
\be
\label{gp}
\gamma^*_L (q)\;\; p (p_N) \to  \pi^+(p_+)\;\; \pi^-(p_-)\;\;
p'(p_{N'})\;.
\ee
%
\begin{figure}[t]
\centerline{\epsfxsize15.0cm\epsffile{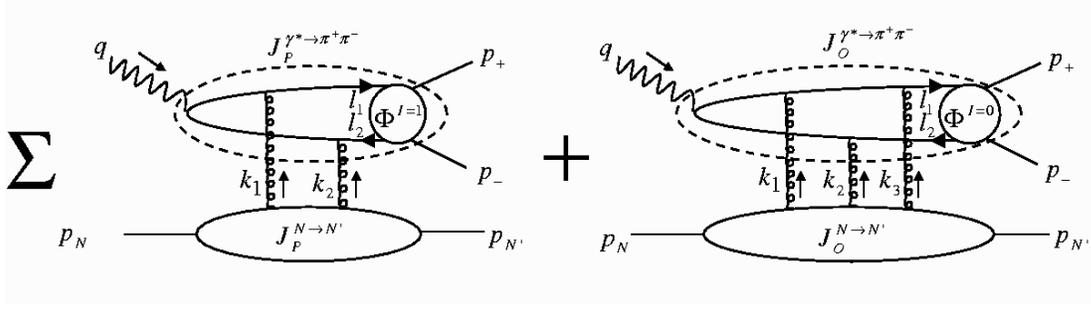}}
\caption[]{\small
Feynman diagrams describing $\pi^+ \pi^-$ electroproduction in 
the Born approximation
 }
\label{fig:2}
\end{figure}
%

We introduce a Sudakov representation with the Sudakov
momenta $p_1,p_2$  obeying 
the equation $s = 2p_1 \cdot p_2$, where $s$ is related to
the total
energy squared of the virtual photon - proton system,
$Q^2$ and the proton target mass $M$ as
$$
(q + p_N)^2 = s - Q^2 +M^2\ \approx s\;,
$$
we get for the virtual photon momentum~:

\be
\label{q}
q^{\mu} = p_1^{\mu} - \frac{Q^2}{s}p_2^{\mu},\;\;\;\;
\ee
and for the momentum of the two pion system~:

\be
\label{2pi}
p_{2\pi}^{\mu} = (1-\frac{\2^2}{s})p_1^{\mu} +\frac{m_{2\pi}^2 +
\2^2 }{s}p_2^{\mu} + p_{2\pi \perp}^{\mu},\;\;\;\;
p_{2\pi \perp}^2=-\2^2\ .
\ee

The  quark ($l_1$) and antiquark ($l_2$) momenta
inside the loop before forming two pion system are parametrized as~:
\begin{equation}
\label{l1}
l_1^{\mu} = z p_1^{\mu} +\frac{m^2+(\vec{l}+z\2)^2}{zs}p_2^{\mu}+
(l_\perp +z p_{2\pi\ \perp})^{\mu}
\end{equation}
\begin{equation}
\label{l2}
l_2^{\mu} = \bar{z} p_1^{\mu} +\frac{m^2+(-\vec{l}+
\bar{z}\2)^2}{\bar{z}s}p_2^{\mu}+
(-l_\perp +\bar{z}\ p_{2\pi\ \perp})^{\mu}
\end{equation}
where $2\vec{l}$ is the relative transverse momentum of the 
quarks forming
the two
pion system and
$\bar{z} = 1-z $, up to small corrections of the order $\2^2/s$. The
collinear approximation of the factorization procedure of the
description of the two pion formation through the generalized
distribution amplitude leads  to the   vector $\vec{l}=\vec{0}$ in
the hard amplitude.

In a similar way as in (\ref{l1}), (\ref{l2}) we parametrize
the momenta of produced pions as
\begin{equation}
\label{p+}
p_+^{\mu} = \zeta p_1^{\mu} +\frac{m_\pi^2+(\p+
\zeta\2)^2}{\zeta s}p_2^{\mu}+
(p_\perp +\zeta p_{2\pi \perp})^{\mu}
\end{equation}
\begin{equation}
\label{p-}
p_-^{\mu} = \bar{\zeta} p_1^{\mu} +\frac{m_\pi^2+(-\p+
\bar{\zeta }\2)^2}{\bar{\zeta }s}p_2^{\mu}+
(-p_\perp +\bar{\zeta }\ p_{2\pi \perp})^{\mu}
\end{equation}
where $2\p$ is now the relative transverse momentum of the produced pions,
$\zeta = \frac{p_2\cdot p_+}{p_2\cdot p_{2\pi}}$ is the fraction of
the longitudinal momentum
$p_{2\pi}$ carried by the produced $\pi^+$, and $\bar{\zeta } = 1-\zeta $.
 The variable $\zeta$ is related to the polar decay angle $\theta$ defined
in the rest frame of the  pion pair by
\be
\label{theta}
\beta \cos \theta = 2\zeta -1\,,\;\;\;
\beta \equiv \sqrt{1 - \frac{4\,m_\pi^2}{m_{2\pi}^2}}\;.
\ee
Since the "longitudinal part" of the two pion wave function depends only
on the angle $\theta$ and doesn't depend on the azimuthal decay angle
$\phi$ (in the same rest frame of the pair) we focus on the calculation
of
forward-backward asymmetries expressed in terms of $\theta$ (see below).

\vskip.1in
The nucleon momentum in the initial state is given by
\begin{equation}
p_{N}^{\mu} = p_2^{\mu} +\frac{M^2}{s}p_1^{\mu}\;.
\end{equation}
The squared momentum transfer  $t=r^2\ $($r^{\mu} = p_{2\pi}^{\mu}
-q^{\mu})$ can be written as
\begin{equation}
t=r^2= -\2^2 + t_{min}, \;\;\;\;t_{min}=
-\frac{M^2(Q^2+m_{2\pi}^2)^2}
{s^2}\;.
\end{equation}

\vskip.2in
{\large \bf 3.~~}
It is well known (see e.g. \cite{Engel} and references therein) that for
large values of $s$ and small momentum transfer $t$ the scattering
amplitudes can be represented as  convolutions over the two-dimensional
transverse momenta of the $t$-channel gluons.

\vskip.1in
For the Pomeron exchange, which corresponds in the Born approximation to
the exchange of two gluons in a colour singlet state,
the impact representation has the form:
\be
\label{pom}
 {\cal M}_P = -i\,s\,\int\;\frac{d^2 \kt_1 \; d^2 \kt_2 \;
  \delta^{(2)}(\kt_1 +\kt_2-\2)}{(2\pi)^2\,\kt_1^2\,\kt_2^2}
 J_P^{\gamma^* \rightarrow \pi^+\pi^-}(\kt_1,\kt_2)\cdot
J_P^{N \rightarrow N'}(\kt_1,\kt_2)
\ee
where $J_P^{\gamma^* \rightarrow \pi^+\pi^-}(\kt_1,\kt_2)$
 and $J_P^{N \rightarrow N'}(\kt_1,\kt_2)$ are the impact factors
 for transition \\
$\gamma^* \to \pi^+\ \pi^-$
via Pomeron exchange 
  and of the nucleon in initial state $N$ into the nucleon in the
 final state $N'$.

\vskip.1in
The corresponding representation for the Odderon exchange, {\em i.e.} the
exchange of three gluons in a colour singlet state, is given by the
formula
\be
\label{odd}
 {\cal M}_O =-\frac{8\,\pi^2\,s}{3!}\int\;\frac{d^2 \kt_1 \; d^2
  \kt_2 d^2 \kt_3\;
  \delta^{(2)}(\kt_1 +\kt_2 +\kt_3-\2)}{(2\pi)^6\,\kt_1^2\,\kt_2^2\,\kt_3^2}
 J_O^{\gamma^* \rightarrow \pi^+\pi^-}\cdot
J_O^{N \rightarrow N'}
\ee
where $J_O^{\gamma^* \rightarrow \pi^+\pi^-}(\kt_1,\kt_2,\kt_3)$
 and $J_O^{N \rightarrow N'}(\kt_1,\kt_2,\kt_2)$ are the impact factors
 for the transition 
$\gamma^* \to \pi^+\ \pi^-$
via Odderon exchange 
  and of the nucleon in initial state $N$ into the nucleon in the
 final state $N'$.

The impact factors are calculated by the standard methods,
see
e.g. Ref. \cite{GI} and references therein. An important aspect of
the present study is the inclusion of an appropriate  two pion
distribution
amplitude which we now discuss.

\vskip.2in
{\large \bf 4.~~}
The two-pion generalized distribution amplitude (GDA) \cite{DGPT, POL}
contains the full strong interactions between the two pions. So far no
experimental information exists on the two-pion GDA. Watson's theorem
imposes that the dynamical
phases of the two-pion GDA are identical to the phase shifts in elastic
$\pi\pi$ scattering as long as $m_{2\pi}$ is below the inelastic
threshold. This relation may be used as an input for a model GDA in the
$m_{2\pi}$-region up to 1~GeV.
For higher values of $m_{2\pi}$ we assume that the phase is still
approximately equal to the $\pi \pi$ phase shift.
 The Odderon exchange process involves the
production of a pion pair in the $C$-even channel which corresponds to
even isospin. In the numerical studies we will  use a
simple ansatz \cite{DGP} for the isosinglet distribution amplitude
$\Phi^{I=0}(z,\zeta,m_{2\pi}^2)$, in a slightly enlarged $m_{2\pi}$ 
range.
We only consider the contributions from $u$- and $d$-quarks, i.e. we
take $n_f~=~2$.

A crucial point  is the choice of the
parametrization of the phases in the
  GDA's. Let us discuss first  the isosinglet $s-$ and $d-$wave phase
shifts, 
$\delta_0$ and $\delta_2$.
Through interference effects, the rapid variation of a phase
shift leads to a characteristic  $m_{2\pi}$-dependence of the asymmetry.
 We use
\bea
&&\Phi^{I=0}(z,\zeta,m_{2\pi}) = 10 z(1-z) (2z-1)\, R_\pi\,\nonumber \\
&& \left[ - \frac{3-\beta^2}{2}\, e^{i\delta_0(m_{2\pi})}\
|BW_{f_0}(m_{2\pi}^2)|
     + \beta^2\ e^{i\delta_2(m_{2\pi})}
\ |BW_{f_2}(m_{2\pi}^2)|\ P_2(\cos \theta)
   \right]\ ,
  \label{model-gda}
\eea
with $R_\pi = 0.5$ and $\beta$ given by Eq. (\ref{theta}).
In our studies we fix the shapes  of the phase
shifts
$\delta_0$ and $\delta_2$
from a
fit to data presented in
\cite{Hyams}.
Let us stress that the exact description of the phase shifts is crucial
for our analysis. It is well known that the Breit-Wigner parametrization
of the amplitude is only a reasonable approximation in the vicinity of the
resonance peaks. In particular the vicinity of the $K\bar K$-threshold
leads to some theoretical uncertainty of our results, although we do not
expect a dramatic effect.  

 $|BW_{f}(m_{2\pi}^2)|$ is the modulus of the Breit-Wigner
amplitudes
\be
\label{f0}
BW_{f_0}(m_{2\pi}^2)= \frac{m_{f_0}^2}{m_{f_0}^2-m_{2\pi}^2-
im_{f_0}\Gamma_{f_0}}\ ,\;\;\;\;
 m_{f_0}=980\,\mbox{MeV}\,,\;\;\;\;\Gamma_{f_0}=50\,- 100\,\mbox{MeV}
\ee
\be
\label{f2}
BW_{f_2}(m_{2\pi}^2)= \frac{m_{f_2}^2}{m_{f_2}^2-m_{2\pi}^2-
im_{f_2}\Gamma_{f_2}}\ ,\;\;\;\;
 m_{f_2}=1275\,\mbox{MeV}\,,\;\;\;\;\Gamma_{f_2}=186\,\mbox{MeV}\ .
\ee
To take into account of the uncertainty in the $f_0$ width, we will
present results for the two extreme allowed values.

\vskip.1in
The Pomeron exchange process involves the
production of a pion pair in the $C$-odd state. Its amplitude can be
fully computed for values of $m_{2\pi}$ where the timelike electromagnetic
pion form factor
$F_\pi(m_{2\pi}^2)$ is known provided the lowest Gegenbauer polynomial
component is dominant. The modulus of $F_\pi$ has been well measured
in the process $e^+ e^- \to \pi^+\pi^-$. By Watson's theorem its phase is
equal to
the $p$-wave phase shift $\delta_1$, provided that $m_{2\pi}$ is in the
range
where $\pi\pi$ scattering is elastic. This is rather well satisfied
for values of 
$m_{2\pi}$ up to 1~GeV. We assume that for $m_{2\pi}$ up to
1.5 GeV
we can still use the phase shifts from $\pi\pi$ elastic
scattering as the phase of the distribution amplitude. 
 In our numerical
studies
we take a  $F_\pi$-parameterization inspired by the $N=1$ model
of~Ref. \cite{Kuehn}
\be
\Phi^{I=1}(z,\zeta,m_{2\pi})
 = 6z(1-z)\beta \cos \theta F_\pi(m_{2\pi}^2) \ ,
\ee
where
\be
\label{FFrho'}
F_\pi(m_{2\pi}^2)= \frac{1}{(1-0.145)}
  BW_\rho\,\frac{1+1.85\cdot 10^{-3}\cdot BW_\omega}{1+1.85\cdot
    10^{-3}}\ ,
\ee
with
\bea
&&BW_\rho(m_{2\pi}^2)= \frac{m_\rho^2}{m_\rho^2-m_{2\pi}^2-
i\sqrt{m_{2\pi}^2}\Gamma_\rho(m_{2\pi}^2)}\ ,  \\
&&\Gamma_\rho(m_{2\pi}^2)= \Gamma_\rho\, \frac{m_\rho^2}{m_{2\pi}^2}\,
\frac{(m_{2\pi}^2-4\,m_\pi^2)^{3/2}}{(m_\rho^2 - 4\,m_\pi^2)^{3/2}}
\ ,\;\;\;
m_\rho=773\,\mbox{MeV}\,, \;\;\;\Gamma_\rho=145\,\mbox{MeV}\ ,\nonumber
\eea
and
\be
BW_{\omega}(m_{2\pi}^2)= \frac{m_\omega^2}{m_\omega^2-m_{2\pi}^2-
im_{\omega}\Gamma_\omega}\ ,\;\;\;\;
 m_{\omega}=782\,\mbox{MeV}\,,\;\;\;\;\Gamma_{\omega}=8.5\,\mbox{MeV}\ .
\ee
The parametrization of the pion form-factor given by Eq.~(\ref{FFrho'})
leads to a reasonable
description of the data on the square of the pion
form-factor, see \cite{Kuehn}. It describes also satisfactorily
 the $p$-wave phase shift in the region of $2\pi$ invariant mass
  smaller than  $1.5 $ GeV. Above $1$ GeV the phase shift given by the
original parametrization of~Ref. \cite{Kuehn} strongly overestimates the
data
points presented in
  \cite{Hyams},  due to a problematic $\rho'(1370)-$ contribution. Since a
correct treatment of phases is crucial for our predictions, we do not
include any  $\rho'-$ contribution.

\vskip.2in
{\large \bf 5.~~}
After choosing  the two pion distribution amplitude, the calculation of the
necessary impact factors is straightforward. Skipping unessential details,
let us now present the final results.
For the longitudinal polarization of virtual photon $\gamma^*_L$ we
obtain for the impact-factor for $\gamma^*_L \rightarrow
\pi^+\,\pi^-$ the following formulas:
\vskip.1in
\begin{itemize}

\item Pomeron exchange:

\be
\label{lP}
J_P^{\gamma^*_L \rightarrow \pi^+\pi^-}(\kt_1,\kt_2) =
-\frac{i\,e\,g^2\,\delta^{ab}\,Q}{2\,N_C}\;
\int\limits_0^1\,dz\,z{\bar z}\,P_P(\kt_1,\kt_2)\,
\Phi^{I=1}(z,\zeta,m_{2\pi}^2)
\ee

\be
P_P(\kt_1,\kt_2)=\frac{1}{z^2\2^2+\mu^2} +\frac{1}{{\bar
    z}^2\2^2+\mu^2}-
\frac{1}{(\kt_1-z\2)^2+\mu^2} -\frac{1}{(\kt_1-{\bar z}\2)^2+\mu^2}
\ee
where $\kt_1+\kt_2=\2$ and $\mu^2 = m_q^2 + z\,{\bar z}\,Q^2$.

\vskip.1in
\item Odderon exchange (where $\kt_1+\kt_2+\kt_3=\2$) :

\be
\label{lO}
J_O^{\gamma^*_L \rightarrow \pi^+\pi^-}(\kt_1,\kt_2,\kt_3) =
-\frac{i\,e\,g^3\,d^{abc}\,Q}{4\,N_C}\;
\int\limits_0^1\,dz\,z{\bar z}\,P_O(\kt_1,\kt_2,\kt_3)\,
\frac{1}{3}\Phi^{I=0}(z,\zeta,m_{2\pi}^2)
\ee

\bea
&&P_O(\kt_1,\kt_2,\kt_3)
=\frac{1}{z^2\2^2+\mu^2} -\frac{1}{{\bar
    z}^2\2^2+\mu^2} \nonumber \\
&&-\sum\limits_{i=1}^3 \left( \frac{1}{(\kt_i-z\2)^2+\mu^2} -
 \frac{1}{(\kt_i-{\bar z}\2)^2+\mu^2}  \right)
\eea
\end{itemize}
The value of the strong coupling constant $g$ in the hard block is assumed
to correspond to the 1-loop running coupling constant with $n_f=2$,
$\alpha_s(Q^2)=\frac{g^2}{4\pi} = {12\pi}/[{29 \ln
(\frac{Q^2}{\Lambda_{QCD}^2})}] $. In order to estimate the theoretical
error we
vary the value of  
$\Lambda_{QCD} = 0.2 \ -\ 0.35\ $GeV. This does 
not lead to any dramatic change in the magnitude of the
asymmetry.

\vskip.1in
Finally we have to fix the soft parts of our amplitudes, i.e.
 the proton impact factors.
They cannot be calculated within perturbation
theory. In our estimates we will use phenomenological, eikonal  models
of these impact factors
proposed in Refs. \cite{protonP} and \cite{protonO}. We take

\begin{itemize}
\item for the Pomeron exchange~:

\be
J_P^{N\rightarrow N'} = i\frac{{\bar g}^2\,\delta^{ab}}{2\,N_C}\,3\,\left[
  \frac{A^2}{A^2+1/2\,\2^2} - \frac{A^2}{A^2
+1/2(\kt_1^2 + \kt_2^2)} \right]\,,\;\;
\ee

\item for the Odderon exchange~:

\be
J_O^{N\rightarrow N'} = -i\frac{{\bar g}^3\,d^{abc}}{4\,N_C}\,3
\left[
 F(\2,0,0) - \sum\limits_{i=1}^3 F(\kt_i,\2-\kt_i,0)
 +2\,F(\kt_1,\kt_2,\kt_3) \right]
\ee
where
\be
F(\kt_1,\kt_2,\kt_3)= \frac{A^2}{A^2 + \frac{1}{2}\left[ (\kt_1-\kt_2)^2 +
(\kt_2-\kt_3)^2 + (\kt_3-\kt_1)^2 \right]  }
\ee
and
$A=\frac{m_\rho}{2}$.
\end{itemize}

In these formula, we have denoted the QCD-coupling constant as ${\bar
g}$.
While it is natural to take $Q^2$ as the scale
of the strong coupling constant in the upper impact factor,
a typical hadronic scale $M^2$ is better suited
 for the  lower one. The value of the coupling constant ${\bar g} =
{ g}(M^2)$  is one of the main sources of theoretical uncertainties of
our numerical results.

In the original Refs. \cite{protonP, protonO} it was assumed that
${\alpha}_{soft} = \frac{\bar g^2}{4\pi} \approx 1$. In view of 
the results of
the recent studies \cite{Dosh}, it seems that this value is too large
for a correct description of the $pp$ differential cross-section in the
region of the dip, and one should  rather take ${\alpha}_{soft} =0.3 -
0.7$.
In order to visualize this rather large uncertainty we
 present
our results with an
error band corresponding to this interval of ${\alpha}_{soft}$. We
 also want
to emphasize that an increase of the value of ${\alpha}_{soft}$ to
${\alpha}_{soft} \approx 1$ raises  our predictions for the asymmetry
by a factor $\approx 1.3$.

\vskip.1in
 {\large \bf 6.~~} Let us now present our
estimates of the charge asymmetry. We define the forward - backward
asymmetry as

\bea
\label{asym}
A(Q^2, t, m_{2\pi}^2)
=\frac{ \int \cos \theta \,d \sigma
(s,Q^2,t,m_{2\pi}^2,\theta)}{
\int d \sigma (s,Q^2,t,m_{2\pi}^2,\theta)}
\nonumber \\
=\frac{ \int\limits_{-1}^{1}\,\cos \theta\,d\cos \theta \,2\ \mbox{Re}
\left[{\cal M}_P^{\gamma^*_L}({\cal M}_O^{\gamma^*_L})^*  \right]}{
\int\limits_{-1}^{1}\,d\cos \theta \left[ |{\cal M}_P^{\gamma^*_L}|^2
+ |{\cal M}_O^{\gamma^*_L}|^2  \right]}
\eea
which may be rewritten as

\be
A(Q^2, t, m_{2\pi}^2) =\frac{\int\limits_{(1-\beta)/2}^{(1+\beta)/2} \,d\zeta
\frac{1}{\beta}(2\zeta
-1)\,2\ \mbox{Re}
\left[{\cal M}_P^{\gamma^*_L}({\cal M}_O^{\gamma^*_L})^*  \right]}{
\int\limits_{(1-\beta)/2}^{(1+\beta)/2}\,d\zeta \left[ |{\cal
M}_P^{\gamma^*_L}|^2
+ |{\cal M}_O^{\gamma^*_L}|^2  \right]     }
\ee

%
\begin{figure}[t]
\centerline{\epsfxsize12.0cm\epsffile{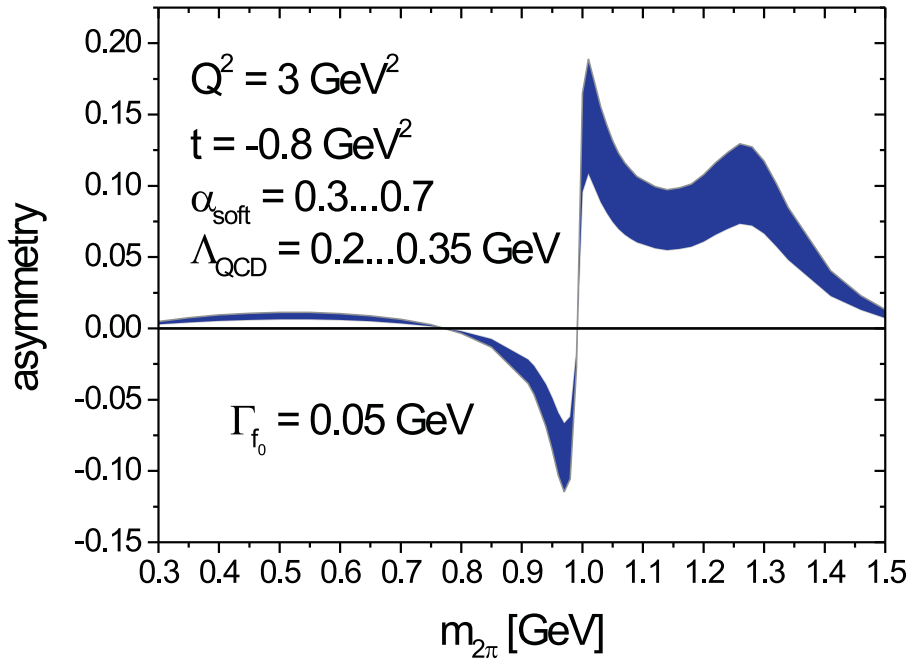}}
\caption[]{\small
Charge asymmetry given by Eq. (\ref{asym}) for a minimal $f_0$ width,
with an error
band showing the uncertainty comming from different values
of $\alpha_{soft}$ and
$\Lambda_{QCD}$.
 }
\label{fig:3}
\end{figure}
%
%
\begin{figure}[t]
\centerline{\epsfxsize12.0cm\epsffile{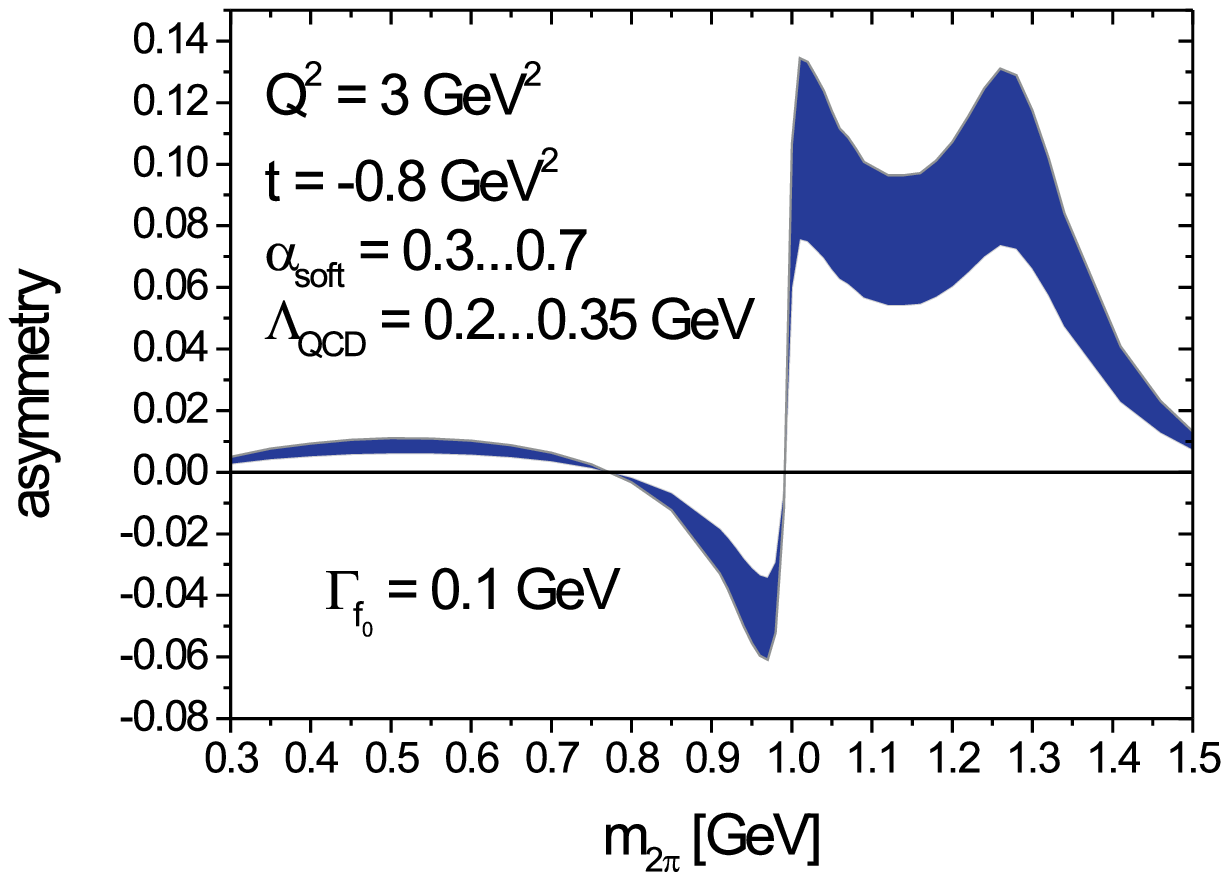}}
\caption[]{\small
Same as Fig. 3 but for maximal $f_0$ width.
 }
\label{fig:4}
\end{figure}
%
Instead of weighted integration of the cross-section it is possible
to perform a full angular analysis. The numerator of the 
asymmetry would then be  
 provided by  the $\cos \theta$-term
 which is characteristic of the longitudinal polarization of the pion
pair, while in the denominator one may extract experimentally 
the contribution of the longitudinally polarized pion pair
in
complete analogy to the case of longitudinally polarized vector mesons.

We checked that the squared Odderon contribution in the denominator can
 be neglected, so that the asymmetry is practically a measure of the
ratio of the Odderon and the Pomeron amplitudes.

There is no $s-$dependence in our framework, within the approximation
which we make,  provided $s$
is large enough for the usual high energy approximation to hold. The
charge asymmetry is plotted in Figs. 3 and 4 as a function of the two
pion invariant mass
$m_{2\pi}$.
%
\begin{figure}[t]
\centerline{\epsfxsize12.0cm\epsffile{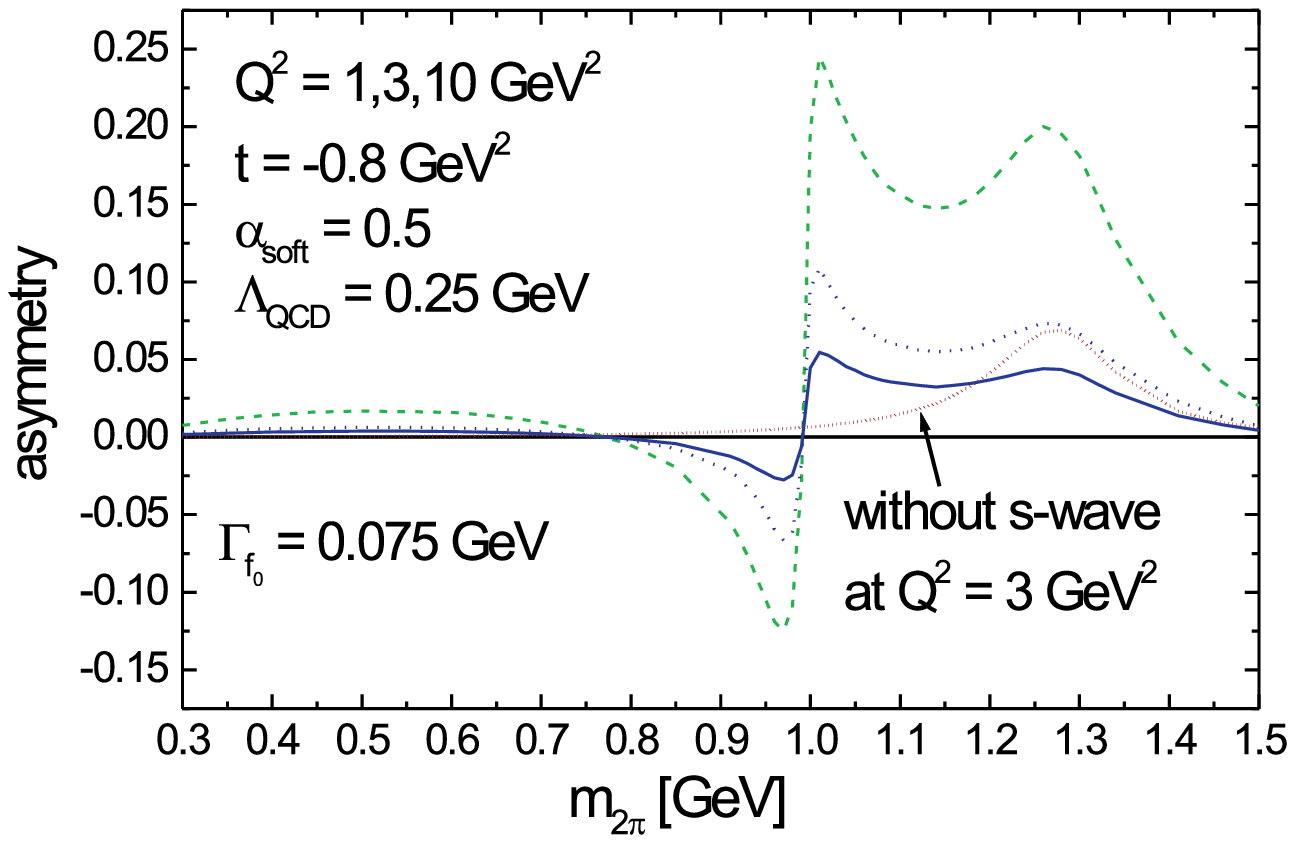}}
\caption[]{\small
$m_{2\pi}$-dependence of the asymmetry for $t = -.8\  $GeV$^2$ and for
different values of $Q^2$: 1 GeV$^2$ (dashed line),  3 GeV$^2$ (dotted
line),  10 GeV$^2$ (solid line); the $d-$wave contribution at
$Q^2=3\ $GeV$^2$ is shown with
a dense dotted line; 
the $f_0$ width has been taken to be  $75\ $MeV.
 }
\label{fig:5}
\end{figure}
%
%
\begin{figure}[t]
\centerline{\epsfxsize12.0cm\epsffile{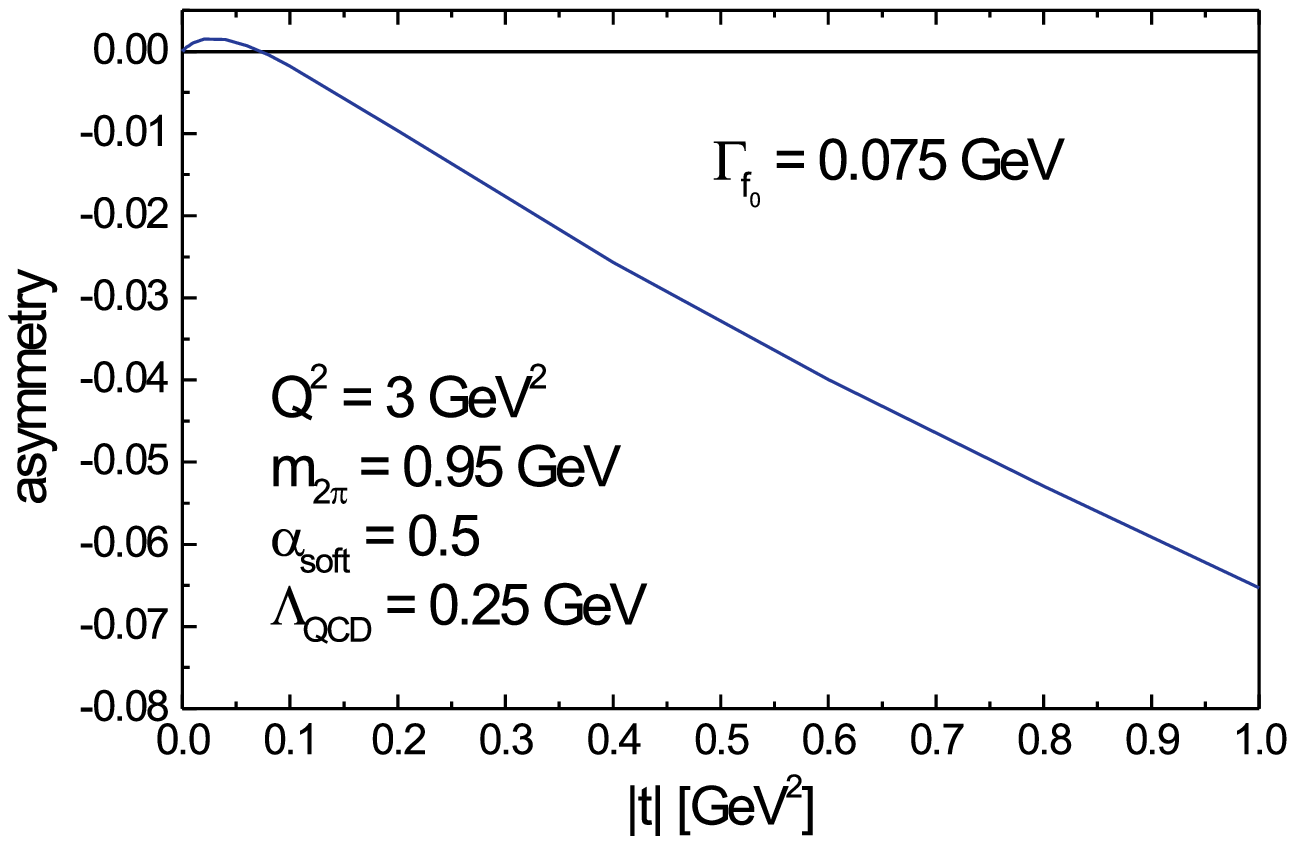}}
\caption[]{\small
$t$-dependence of the asymmetry for $Q^2 = 3\ $GeV$^2$ and
$m_{2\pi}=0.95\ $GeV; the $f_0$
width has been taken to be $75\ $MeV.
 }
\label{fig:6}
\end{figure}
%
The main characteristic is the high numerical value that we
get for values of $m_{2\pi}$ around the $f_0$ mass. This is in strong
contrast
to
the result obtained in a completely different framework by the authors
of Ref. {\cite{Nikolaev}}. The characterisic $m_{2\pi}$ dependence 
in Fig. 3 and Fig. 4
 is completely understood in terms of the $\pi \pi$ phase shifts and the
factor
$\sin (\delta^0-\delta^1)$.
The phase difference vanishes for $m_{2\pi} \approx 0.75\ $GeV and 
$m_{2\pi} \approx 1\ $GeV resulting in two zeros of the charge asymmetry.
The magnitude of the asymmetry  depends much on
the width of the $f_0$ meson which is estimated to be $50-100$ MeV. The
curve shown in the Fig. 3 and in the Fig. 4  are obtained with the extreme
values of this width. 
The $d-$wave contribution shows up around 1.3 GeV on Figs. 3, 4 and 5. 
In Fig. 5 we show also the
$Q^2$ dependence of the asymmetry which turns out to be rather moderate.

The $t$ dependence of the asymmetry, which is plotted in Fig. 6, is quite
interesting
since it has a characteristic zero around $t = -0.1$ GeV$^2$.
This zero in
the odderon amplitude has already been discussed in Ref. 
\cite{Vacca2}. 
The practical outcome of this $t$-dependence is that it is better to
focus on moderate values
of t and, if an integration over t is needed in order to improve the
statistics, to
choose at least $-t\approx 0.1\ $GeV$^{2}$ as a lower integration limit.


\vskip.1in
{\large \bf 7.~~} Let us now comment on the possible theoretical
uncertainties of our calculations.

\begin{itemize}

\item Higher twist contributions and corrections to the calculated
expressions of order $m_{2\pi}^2/Q^2$ or $t/Q^2$  may
well be non negligible. For instance, $k_\perp$ effects in the $\pi^+\
\pi^-$
wave function mentioned above may be important at relatively low $Q^2$
\cite{FKS} as well 
as the effects of transversely polarized photon. An
estimate of these corrections
is certainly desirable but  clearly out of the scope of the present paper.

\item QCD evolution {\`a} la BFKL and $\ln(s/\Lambda_{QCD} ^2)$
corrections
may
be calculated for the Pomeron and Odderon exchanges as well as the
 $\ln(Q^2/\Lambda_{QCD} ^2)$ corrections for the generalized distribution
amplitudes. We expect that they do not have drastic effects on
ratios such as the charge asymmetry which we have calculated. This
may be
controlled in the future.

\item  The exact values of the scales of the coupling constants 
should be
determined by the
presently unknown higher order corrections in the hard and soft parts,
respectively.
It is interesting that in the  case of the dominance of 
Abelian contributions to  radiative corrections,
the coupling renormalization  would be reduced, by use of the Ward
identities,  to bubble insertions into the gluon line, so that
the scales would be common for the hard and soft parts and determined by
the  gluon off-shellness. Due to the  integration over the gluon
momenta the latter
should be averaged to some intermediate value between $Q^2$ and $M^2$,
so that the above naive estimate may still be valid.

\item  The $\pi \pi$ distribution amplitude is a non perturbative object
which we certainly do not exactly know. Although its phase is
theoretically under control in the lower mass range, its
magnitude and its $z-$ and $\zeta -$ dependence may be quite different
from the simple ansatz that we have adopted. Let us stress however that
parts of this uncertainty can be resolved by other experiments, namely two
pion production in $e \gamma $ collisions \cite{DGP} and in
$ep$ collisions at medium energies \cite{LDSPG}
sensitive to the C-even and C-odd components of GDA.

\end{itemize}

Let us now briefly indicate  possible future studies closely related to
the approach that we have developed.

\begin{itemize}

\item Single spin asymmetries should show up in the same reaction 
\cite{todd}.
Longitudinally polarized electron beams are providing  circularly
polarized virtual photons. In turn, they give rise to  single spin
azimuthal  asymmetries, similar to those appearing in the $\gamma^*
\gamma$ production of pion  pairs \cite{DGP} and triplets \cite{PT}. The
expression for the asymmetry  is similar to the one for the charge
asymmetry with the notable difference that the imaginary, instead of
real part of the relevant product of amplitudes appears in the numerator
of the analog of
Eq.\ref{asym}. The consequence is that, in the case of the 
Pomeron-Odderon
interference, whose amplitudes are mostly imaginary and real,
respectively, there appears a factor
$\cos(\delta_0-\delta_1)$ instead of $\sin (\delta_0-\delta_1)$, which
leads to a completely different $m_{2\pi}$ dependence.
Note that in the case of the calculations of the single spin asymmetry one
has to take into account transverse polarizations of a photon.

\item Charge asymmetries can also be studied in the interesting case
of $\gamma^* \gamma^*$ scattering which may be measured at $e^+ e^-$
colliders. A specific feature of this case is that both impact factors
are calculable within perturbative QCD.

\end{itemize}

\vskip.1in

{\large \bf 8.~~} In conclusion, let us stress  that we have
demonstrated that the
understanding of diffractive processes within perturbative QCD is bound to the
discovery of sizable charge asymmetries in electroproduction of two charged
mesons.

We applied the powerful tool of QCD factorization when the hard part
(subprocess) is calculated perturbatively, while the soft ingredients
(GDA and proton impact factor) should be modeled or, better, measured,  
which poses a new challenging problem for experimentalists, 

Data on this reaction in the kinematical domain suitable for our
calculation (i.e.
large $s$, small $t$, $Q^2$ above $1\ $GeV$^2$ and $m_{2\pi}$ below $1.5\
$ GeV)
should be easy to get and analyze by the experimental set-ups H1
 \cite{Olsson} and ZEUS \cite{ZEUS} at HERA. We are eagerly waiting
for this
confrontation of theory
with data, which should lead us towards the discovery  of the Odderon.

\vskip.1in
{\large \bf 9.~~}{\bf Acknowledgements}

\vskip.1in
We acknowledge useful discussions with S.~Brodsky, A.V.~Efremov, 
I.F.~Ginzburg, 
B.~Loiseau,
S.V.~Mikhailov, B.~Nicolescu, J.~Olsson, T.N.~Pham, A.~Sch{\"a}fer, 
T.N.~Truong and S.~Wallon.

This work is supported in part
by the TMR and IHRP Programmes of the European Union, Contracts
No.~FMRX-CT98-0194 and No.~HPRN-CT-2000-00130, RFFI Grant 16696,
and iNTAS Project 587 (Call 2000).


\begin{thebibliography}{99}

\bibitem{LN} L.~Lukaszuk and B.~Nicolescu, Lett. Nuovo Cim. {\bf 
8} (1973) 405


\bibitem{BKP} J.~Bartels, Nucl.\ Phys.\ B {\bf 151} (1979) 293; {\it ibid}
B {\bf 175} (1980) 365; \\
J.~Kwiecinski and M.~Praszalowicz,
Phys.\ Lett.\ B {\bf 94} (1980) 413.

\bibitem{Levodd} L.N. Lipatov, Phys.\
  Lett.\ B {\bf 309} (1993) 394~; JETP Lett. {\bf 59} (1994) 596~; Sov.\
Phys.\ JETP\
Lett.\ {\bf 59} (1994) 571; \\
L.D.~Fadeev and G.P. Korchemsky, Phys.\ Lett.\ B {\bf 342} (1995) 311

\bibitem{JW} R.A.~Janik and J.~Wosiek, Phys. Rev. Lett. {\bf 82} 
(1999) 1092

\bibitem{Vacca1} J.~Bartels, L.N.~Lipatov and G.P.~Vacca, Phys. Lett.
B {\bf 477} (2000) 178


\bibitem{Korch} 
G.~P.~Korchemsky, J.~Kotanski and A.~N.~Manashov,
hep-ph/0111185.

\bibitem{BFKL} E.A.~Kuraev, L.N.~Lipatov and V.S.~Fadin, Sov. JETP
{\bf 44} (1976) 443~; {\it ibid} {\bf 45} (1977) 199~; Ya.Ya.~Balitsky and
L.N.~Lipatov, Sov.\ J.\ Nucl.\ Phys.\ {\bf 28} (1978) 822~;
L.N.~Lipatov, Sov. Phys. JETP {\bf 63} (1986) 904




\bibitem{KM} J.~Czyzewski, J.~Kwiecinski, L.~Motyka and
  M.~Sadzikowski, Phys. Lett. B {\bf 398} (1997) 400~; erratum {\it ibid}
B {\bf 411} (1997) 402


\bibitem{Engel} R.~Engel, D.Yu.~Ivanov, R.~Kirschner and
  L.~Szymanowski, Eur. Phys. J. C {\bf 4} (1998) 93

\bibitem{Vacca2} J.~Bartels, M.A.~Braun, D.~Colferai and G.P.~Vacca,
  Eur. Phys. J. C {\bf 20} (2001) 323



\bibitem{Olsson} J.~Olsson (for the H1 Collab.), hep-ex/0112012

\bibitem{Dosh} E.R.~Berger, A.~Donnachie, H.G.~Dosch, W.~Kilian,
  O.~Nachtmann and M.~Reuter, Eur. Phys. J. C {\bf 9} (1999) 491

\bibitem{Doshrecent} H.G.~Dosch, C.~Ewerz and V.~Schatz, hep-ph/0201294


\bibitem{Brodsky} S.J.~Brodsky, J.~Rathsman and C.~Merino, Phys. Lett.
B {\bf 461} (1999) 114

\bibitem{Nikolaev} I.P.~Ivanov, N.N.~Nikolaev and I.F. Ginzburg,
hep-ph/0110181


\bibitem{sigmaLT} M. Derrick {\it et al}, Phys.\ Lett.\  B{\bf 356}
(1995) 601; P. Amaudruz {\it et al}, Z. Phys. C{\bf 51} (1991) 387; 
M. Arneodo {\it et al}, Nucl. Phys. B{\bf 429} (1994) 503 

\bibitem{DGPT}
M.~Diehl, T.~Gousset, B.~Pire and O.V.~Teryaev,
Phys.\ Rev.\ Lett.\  {\bf 81} (1998) 1782


\bibitem{GI} I.F.~Ginzburg and D.Yu.~Ivanov, Nucl. Phys. B {\bf
388} (1992) 376


\bibitem{POL}
M.~V.~Polyakov and C.~Weiss,
Phys.\ Rev.\ D {\bf 59} (1999) 091502 
[arXiv:hep-ph/9806390];

M.~V.~Polyakov,
Nucl.\ Phys.\  B {\bf 555} (1999) 231



\bibitem{DGP}
M.~Diehl, T.~Gousset and B.~Pire,
Phys.\ Rev.\ D {\bf 62} (2000) 073014



\bibitem{Hyams} B.~Hyams {\it et al.}, Nucl.\ Phys.\ B {\bf 64} (1973)
134  \\
D.V.~Bugg, B.S.~Zou and A.V.~Sarantsev, Nucl.\ Phys.\ B{\bf 471} (1996) 59
 \\
R.~Kaminski, L.~Lesniak and K.~Rybicki, 
Acta Phys.Polon. B {\bf 31} (2000) 895 

\bibitem{Kuehn} J.H. K{\"u}hn and A. Santamaria, Z.\ Phys.\ C {\bf 48}
(1990) 445 


\bibitem{protonP} J.F.~Gunion and D.E.~Soper, Phys. Rev. D {\bf
15} (1977) 2617

\bibitem{protonO} M.~Fukugita and J.~Kwiecinski, Phys. Lett. B
{\bf 83} (1979) 119

\bibitem{FKS} L. Frankfurt, W. Koepf and M. Strikman, Phys. Rev. D {\bf
54} (1996) 3194

\bibitem{LDSPG}
B.~Lehmann-Dronke, P.~V.~Pobylitsa, M.~V.~Polyakov, A.~Sch{\"a}fer and
K.~Goeke,
Phys.\ Lett.\ B {\bf 475} (2000) 147 
[arXiv:hep-ph/9910310];

B.~Lehmann-Dronke, A.~Sch{\"a}fer, M.~V.~Polyakov and K.~Goeke,
Phys.\ Rev.\ D {\bf 63} (2001) 114001 
[arXiv:hep-ph/0012108].

\bibitem{todd} O.V. Teryaev, {\it T-odd fracture functions},
in "SPIN-01", Proceedings of International Workshop, Dubna, 2002.

\bibitem{PT}
B.~Pire and O.~V.~Teryaev,
Phys.\ Lett.\ B {\bf 496} (2000) 76 

\bibitem{ZEUS}
J.~Breitweg {\it et al.}  [ZEUS Collaboration],
Eur.\ Phys.\ J.\ C {\bf 6} (1999) 603
[arXiv:hep-ex/9808020].




\end{thebibliography}
\end{document}